\documentclass[manuscript]{aastex}
\usepackage{aas_macros,url,natbib,amssymb,amsmath,CJK,lineno}
\usepackage[utf8x]{inputenc}

\shorttitle{Bangs and Meteors from the Quiet Comet 15P/Finlay}
\shortauthors{Ye et al.}

\begin{document}
\begin{CJK*}{UTF8}{gbsn}

\title{Bangs and Meteors from the Quiet Comet 15P/Finlay}

\author{Quan-Zhi Ye (叶泉志)}
\affil{Department of Physics and Astronomy, The University of Western Ontario, London, Ontario N6A 3K7, Canada}
\email{qye22@uwo.ca}

\author{Peter G. Brown}
\affil{Department of Physics and Astronomy, The University of Western Ontario, London, Ontario N6A 3K7, Canada}
\affil{Centre for Planetary Science and Exploration, The University of Western Ontario, London, Ontario N6A 5B8, Canada}

\author{Charles Bell}
\affil{Vicksburg, MS, U.S.A.}

\author{Xing Gao (高兴)}
\affil{No. 1 Senior High School of \"{U}rumqi, \"{U}rumqi, Xinjiang, China}

\author{Martin Ma\v{s}ek}
\affil{Institute of Physics Czech Academy of Sciences, Na Slovance 1999/2, 182 21, Praha, Czech Republic}

\and

\author{Man-To Hui (许文韬)}
\affil{Department of Earth, Planetary and Space Sciences, University of California at Los Angeles, 595 Charles Young Drive East, Los Angeles, CA 90095-1567, U.S.A.}

\begin{abstract}
Jupiter-family comet 15P/Finlay has been reportedly quiet in activity for over a century but has harbored two outbursts during its 2014/2015 perihelion passage. Here we present an analysis of these two outbursts using a set of cometary observations. The outbursts took place between 2014 Dec. 15.4--16.0 UT and 2015 Jan. 15.5--16.0 UT as constrained by ground-based and spacecraft observations. We find a characteristic ejection speed of $V_0=300$ to $650 \mathrm{m \cdot s^{-1}}$ for the ejecta of the first outburst and $V_0=550$ to $750 \mathrm{m \cdot s^{-1}}$ for that of the second outburst using a Monte Carlo dust model. The mass of the ejecta is calculated to be $M_\mathrm{d}=2$ to $3\times10^5 \mathrm{kg}$ for the first outburst and $M_\mathrm{d}=4$ to $5\times10^5 \mathrm{kg}$ for the second outburst, corresponds to less than $10^{-7}$ of the nucleus mass. The specific energy of the two outbursts is found to be $0.3$ to $2\times10^5 \mathrm{J \cdot kg^{-1}}$. We also revisit the long-standing puzzle of the non-detection of the hypothetical Finlayid meteor shower by performing a cued search using the 13-year data from the Canadian Meteor Orbit Radar, which does not reveal any positives. The Earth will pass the 2014/2015 outburst ejecta around 2021 Oct. 6 at 22 h UT to Oct. 7 at 1 h UT, with a chance for some significant meteor activity in the radio range, which may provide further clues to the Finlayid puzzle. A southerly radiant in the constellation of Ara will favor the observers in the southern tip of Africa.
\end{abstract}


\section{Introduction}
\label{sec:intro}

Small bodies in the inner solar system are historically classified based on their appearance. Comet refers to an object with extended appearance and sometimes one or several tails; while asteroid refers to an object that is much smaller than the major planets and appears star-like. Classification by orbital dynamics of these bodies shows that the dynamical characteristics of the two groups of objects differ as well: comets usually possess highly elliptical, parabolic or hyperbolic orbits, while asteroids usually possess more circular orbits.

Bodies which deviate from these trends (i.e. comets in asteroidal orbits, or asteorids in cometary orbits) are of significant interest, as their dynamical evolution and/or physical properties are apparently exceptional. Although the first such outlier was officially recognized no later than 1989 \citep[e.g. the case of 95P/(2060) Chiron, c.f.][]{1989IAUC.4770....1M}, most outliers were not found until recently with the commissioning of a number of near-Earth asteroid searching/follow-up programs.  Due to their distinct appearance, comet-like objects in asteroid-like orbits, or ``active asteroids'' \citep{2012AJ....143...66J}, are more straightforward to recognize due to their significant morphological change during the transition to comet-like state, and so the recognition is usually robust. In contrast,  their counterparts, asteroid-like objects in cometary orbits (ACOs), due to their nature, are considerably more difficult to identify. From an orbital perspective, ACOs are most easily interpreted to be comets that have exhausted their volatiles (or have their volatiles permanently buried by their crusts) so that they appear asteroidal, i.e. they become ``dormant comets''. However, dynamical studies have shown that a significant fraction of ACOs could be asteroids leaking out from the main-belt that are temporarily residing in comet-like orbits \citet{2014Icar..234...66T, Fernandez2015c}, therefore complicating the effort on disentangle orbital properties from physical properties of these bodies.

One approach to identify dormant comets in the ACO population is to look at comets at an intermediate state between active comets and dormant comets, sometimes labeled as comet-asteroid transition objects \citep[CATOs; e.g.][]{Licandro2007}. A handful of such objects have been suggested, such as 107P/(4015) Wilson-Harrington \citep{Ishiguro2011}, 209P/LINEAR \citep{2014MNRAS.437.3283Y} and (3552) Don Quixote \citep{2014ApJ...781...25M}. However, these objects are usually faint and produce little dust, presenting a challenge for further investigation of their surface and dust properties.

15P/Finlay, a Jupiter-family comet (JFC), has been reportedly faint and tail-less since its discovery in 1886 \citep{2004come.book.....K, 2008come.book.....K, 2010come.book.....K}. The comet has a small Minimum Orbit Intersection Distance (MOID) of 0.0094~AU to the Earth's orbit, but has never been associated with any known meteor shower \citep{1999MNRAS.310..168B}. Coupled with the fact that 15P/Finlay has shown a systematic decrease of maximum brightness at each perihelion passage in the past century it has been suggested that the comet is approaching a state of complete dormancy \citep{1989BAICz..40..269K}.

However, during its current perihelion passage, 15P/Finlay exhibited two outbursts, each producing a parabolic ``shell'' around the original coma accompanied by a straight, freshly-formed ``tail'' in the anti-sunward direction. This resembles the historic outburst of 17P/Holmes in 2007 \citep{2007IAUC.8886....1B} albeit at a much smaller scale. However, it is notable that 17P/Holmes's outbursts took place at a larger heliocentric distance (2.4~AU) than those of 15P/Finlay ($\sim1.0$~AU), therefore the underlying mechanism may not be necessarily the same although the similarity of their overall appearances is striking.

The outbursts of 15P/Finlay are significant in another context: as an Earth-approaching comet, the outburst ejecta may find their way to the Earth, creating a meteor outburst. Previously, numerical simulation by Mikhail Maslov\footnote{\url{http://feraj.narod.ru/Radiants/Predictions/1901-2100eng/Finlayids1901-2100predeng.html}, accessed 2015 Jan. 17.} has suggested that the material released in 2014 will have a direct encounter with the Earth in 2021, which may produce a meteor outburst with Zenith Hourly Rate (ZHR) up to 50. Recent calculation by Mikiya Sato\footnote{\url{https://groups.yahoo.com/neo/groups/meteorobs/conversations/messages/44030}, retrieved 2015 Mar. 4.} also arrived at similar results. An outburst from the parent comet may result in a stronger meteor event depending on the ejection velocity and planetary perturbation. Potential meteor observations allow us to directly sample materials from a dormant comet candidate without a dedicated space mission, which may help in understanding the comet itself as well as the dormant comets as a population. In particular, meteor activity from ACOs can help establish prior periods of activity and constrain the dust production history of ACOs.

In this work, we present an analysis of the observations of 15P/Finlay taken during the two 2014/2015 outbursts. The goal is to understand the underlying nature of the outburst as well as the evolutionary status of the comet. We also examine the yet-to-be-discovered Finlayid meteor shower and especially the potential 2021 meteor outburst. Non-detection of the shower places constraints on past dust production history of 15P/Finlay.

\section{Observations}

\subsection{Amalgamation of Outburst Reports}
\label{sec:obs-reports}

\begin{figure*}
\includegraphics[width=\textwidth]{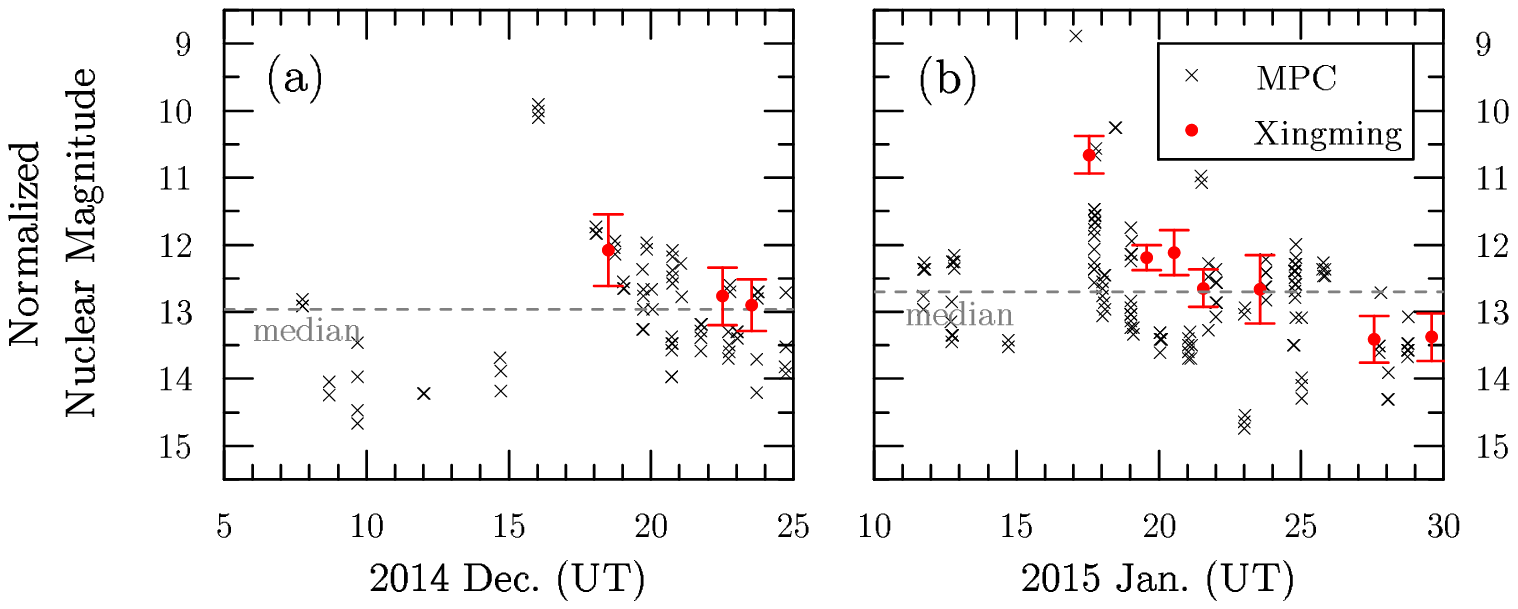}
\caption{Nucleus magnitude of 15P/Finlay around the time of (a) the first outburst, and (b) the second outburst. The Minor Planet Center (MPC) magnitudes (plotted in crosses) are extracted from the Observations Database on the MPC website. The Xingming magnitudes (plotted in red dots) are derived from the monitoring observations by the Xingming 0.35-m telescope with aperture radius $\rho=5000$~km. The magnitudes are normalized to $\varDelta=r_\mathrm{h}=1$~AU assuming a brightening rate $n=4$ \citep{Everhart1967}.}
\label{fig:report}
\end{figure*}

\begin{figure*}
\includegraphics[width=0.85\textwidth]{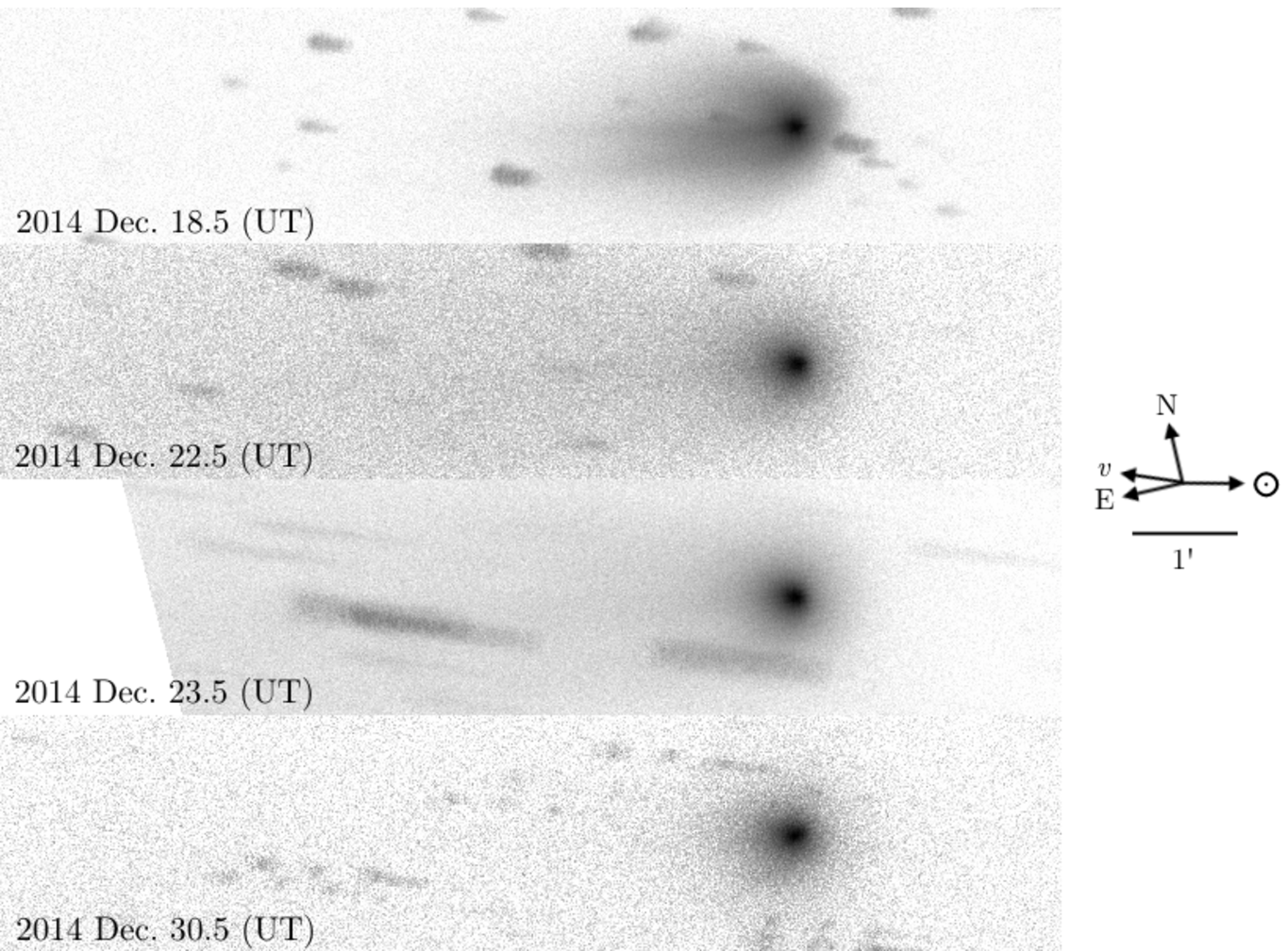}
\caption{Composite images of 15P/Finlay for the first outburst as observed at Xingming Observatory. The images have been stretched in asinh scale. The scale bar shows the direction to the Sun, the comet's velocity vector and the directions of the plane of the sky.}
\label{fig:morp1}
\end{figure*}

The first outburst of 15P/Finlay took place in the late hours of 2014 Dec. 15, the timing being constrained by reports from Christopher Wyatt (Walcha, Australia; Dec. 15.43 UT) and Slooh.com Chile Observatory (La Dehesa, Chile; Dec. 16.04 UT)\footnote{Wyatt's observation is accessible through the Comet Observation Database (COBS), available at \url{http://www.cobs.si}; observation from Slooh.com Chile Observatory was published in the Minor Planet Circular No. 90932 (\url{http://www.minorplanetcenter.net/iau/ECS/MPCArchive/2015/MPC_20150105.pdf}). Both resources were accessed on 2015 Jan. 17.}. During the outburst, the comet brightened by about 3 magnitudes and developed a spiky tail. The tail diluted into the background with the brightness returning to its normal range by Dec. 21--22 (Figure~\ref{fig:report}a, \ref{fig:morp1}).

\begin{figure*}
\includegraphics[width=0.85\textwidth]{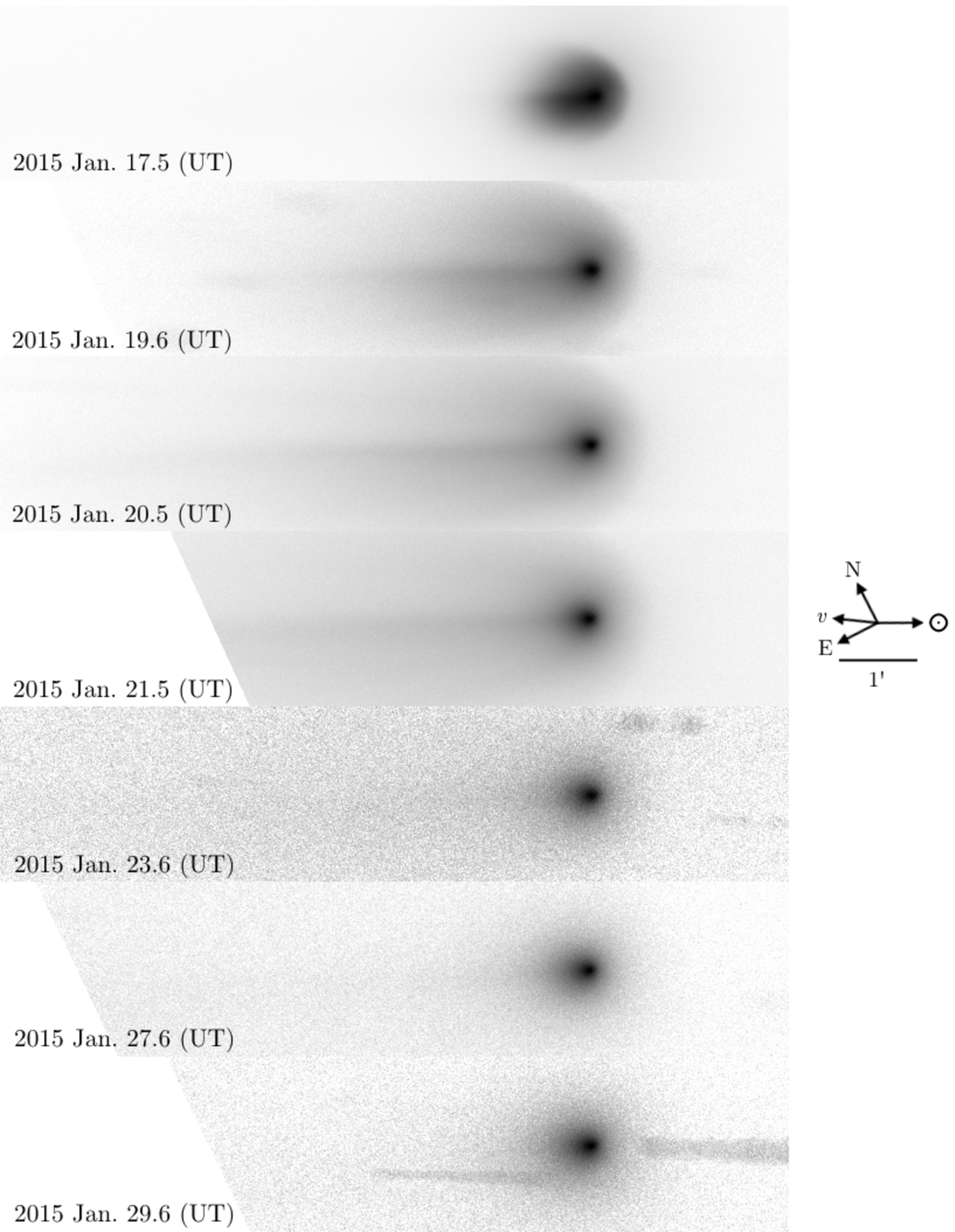}
\caption{Composite images of 15P/Finlay images for the second outburst as observed at Xingming Observatory. The images have been stretched in asinh scale. The scale bar shows the direction to the Sun, the comet's velocity vector and the directions of the plane of the sky.}
\label{fig:morp2}
\end{figure*}

The second outburst took place around 2015 Jan. 16.0 UT as noted by Alan Hale (Cloudcroft, NM) at Jan. 16.07 who noted ``very bright, almost star-like central condensation'' that was absent in the earlier observations\footnote{\url{https://groups.yahoo.com/neo/groups/comets-ml/conversations/messages/24322}, accessed on 2015 Jul. 17, as well as private communication with Alan Hale.}. The last negative observation comes from the Solar Wind ANisotropies all-sky hydrogen Ly-$\alpha$ camera (SWAN) on-board the Solar and Heliospheric Observatory (SOHO) \citep{Bertaux1995a} around Jan. 15.5 UT\footnote{\url{http://sohowww.nascom.nasa.gov/data/summary/swan/swan-images.html}), accessed on 2015 Feb. 9.}. The outburst was subsequently noted independently by Guo Zheng-Qiang (Shenyang, China; Jan. 16.43 UT) and Michael Mattiazzo (Swan Hill, Australia; Jan. 16.45 UT)\footnote{Guo's observation was posted on \url{http://www.astronomy.com.cn/bbs/thread-305185-1-1.html} (in Chinese), Mattiazzo's report was posted on his Facebook page. Both resources were accessed on 2015 Jan. 17.}, as well as on the SWAN image taken near Jan. 16.5 UT (next to the last non-detection image). The comet brightened by about 4--5 magnitudes during the second outburst, again with a freshly-formed tail. The brightness returned to the normal range around Jan. 20 (Figure~\ref{fig:report}b), but the tail lingered for a few more days until around Jan. 30 (Figure~\ref{fig:morp2}).

\subsection{Observation and Image Process}
\label{sec:obs-morphological}

After receiving the reports of the outbursts, 15P/Finlay was monitored using the facilities at F(/Ph)otometric Robotic Atmospheric Monitor (FRAM) located at Pierre Auger Observatory (Argentina), Xingming Observatory (China) and Vicksburg (U.S.). The FRAM observations were conducted with a 0.3-m f/6.6 telescope equipped with a Kodak KAF-1603ME sensor, which gives a resolution of $0.93''$; the Xingming observations were conducted with a 0.35-m f/6.9 telescope with a Kodak KAF-8300 sensor, images are binned by 2, which gives a resolution of $1.2''$. The Vicksburg observations were conducted with a 0.3-m f/10 telescope equipped with Kodak KAF-1600 sensor which gives a resolution of $1.87''$.

The observations at FRAM and Vicksburg are intended for dust modeling, as such they were conducted with a Cousins $R$ filter that blocks flux from major cometary gaseous emissions (such as CN, C$_2$, C$_3$). The Xingming observations were conducted with wider temporal coverage but without a filter, intended as a continuous monitor of the development of the outburst. Details of the observations are summarized in Table~\ref{tbl:obs-summary}. The images are processed using standard procedure (bias subtraction, dark subtraction, flat division), with plate constants solved using UCAC~4 catalog \citep{Zacharias2013}. The images are then medianly combined following the motion of the comet.

The composite images from FRAM and Vicksburg are collapsed into a 1-dimensional profile. This is necessary as the considerable irregularities of the near-nucleus dust (i.e. localized jets) complicate the modeling work. The orbital plane angle at the two outbursts were also shallow enough ($\sim 4^\circ$) to minimize the information loss during the image collapse. The 1-dimensional profile is simply derived from averaging a $2'$ wide strip along the Sun-comet axis, with the width of $2'$ corresponding to the maximum width of the tail.

\section{Analysis}

\subsection{General Morphology and Evolution of the Outbursts}
\label{sec:morphology}

The composite images from the monitoring observations at Xingming (Figure~\ref{fig:morp1} and~\ref{fig:morp2}) show that the morphologies and evolution of both outbursts are comparable: both outbursts produced a newly-formed dust shell that is slightly asymmetric with respect to the comet-Sun axis; the dust shell expands as time goes by and fades into the background within $\sim 1$ week.

We perform aperture photometry with the Xingming data. This is motivated by the considerable scatter shown on the magnitudes provided by the Minor Planet Center (MPC)\footnote{Available from the MPC Observations Database, \url{http://www.minorplanetcenter.net/db_search}, retrieved 2015 Feb. 3.}, possibly due to different instrumental and measurement settings adopted by different observers. Data from FRAM and Vicksburg is not used at this stage to avoid the complication due to instrumental differences. We use an aperture of $\rho=5000$~km as projected at the distance of the comet centered at the nucleus. Both 0.35-m and MPC magnitudes are reduced to ``normalized'' magnitudes at $r_\mathrm{H}=\varDelta=1$~AU using $M_\mathrm{N}=m_\mathrm{N} - 5\log{\varDelta} - 2.5n\log{r_\mathrm{H}}$, where $M_\mathrm{N}$ and $m_\mathrm{N}$ are normalized and observed nuclear magnitudes, $r_\mathrm{H}$ and $\varDelta$ are heliocentric and geocentric distances in AU, and $n=4$ is the canonical brightening rate exponent \citep{Everhart1967}. The photometric calibration is performed using the V-band data from the AAVSO All-Sky Photometric Survey (APASS) catalog \citep{Henden2012} as the Xingming system is most sensitive at V-band. As shown in Figure~\ref{fig:report}, the characteristic outburst decay time (i.e. the time elapsed from the peak of the outburst to the point that the brightness reaches $1/e$ of the peak brightness) is estimated to be at the order of 1~d.

\subsection{Dust Model and Kinematics of the Ejecta}
\label{sec:kinematics}

To understand the dust produced by the outburst event, we model the observations using a Monte Carlo dust model developed in our earlier works \citep[e.g.][]{2014ApJ...787..115Y, Ye2016}.

The dynamical evolution of the cometary dust is controlled by the ratio between radiation pressure and solar gravity, $\beta_\mathrm{rp}=5.7\times10^{-4}/(\rho_\mathrm{d} a_\mathrm{d})$, where $\rho_\mathrm{d}$ is the bulk density of the dust and $a_\mathrm{d}$ the diameter of the dust, both in SI units \citep{Wyatt1950c}, as well as the initial ejection velocity of the dust. The latter is defined as

\begin{equation}
v_\mathrm{ej} = V_0 \beta_\mathrm{rp}^{1/2} \cdot \nu
\end{equation}

\noindent where $V_0$ is the mean ejection speed of a dust particle of $\beta_\mathrm{rp}=1$ and $\nu$ follows a Gaussian probability density function:

\begin{equation}
P(\nu) = \mathcal{N}(1,\sigma_\nu^2)
\end{equation}

\noindent where $\sigma_\nu$ is the standard deviation of $\nu$, used to account for the physical spread $\nu$ due to the shape of the dusts. In this work we use $\sigma_\nu=0.3$ following exploration by, e.g. \citet{Ishiguro2014, Jewitt2014} and \citet{Ye2016}.

We assume the dust size follows a simple power-law with a differential size index of $q$, and the that observed flux is solely contributed by scattered light from the dust particles. Hence, the dust production rate is expressed as

\begin{equation}
N(r_\mathrm{H}, a_\mathrm{d}) \mathrm{d} a_\mathrm{d} = N_0 \left( \frac{a_\mathrm{d}}{1~\micron} \right)^{-q} \mathrm{d} a_\mathrm{d}
\end{equation}

\noindent where $N_0$ is the mean dust production rate of $1~\micron$ particles.

Simulated particles are symmetrically released from the nucleus. For both outbursts, two possible outburst epochs are tested, each correspond to either the epochs of the last negative (non-outburst) report or the first positive report. For the first outburst, outburst epochs of 2014 Dec. 15.4 UT (as indicated by Wyatt's negative report) and 16.0 UT (indicated by Slooh.com's positive report) are tested; for the second outburst, outburst epochs of 2015 Jan. 15.5 UT (indicated by SOHO/SWAN's negative report) and 16.0 UT (indicated by Hale's positive report) are tested. The production rate peaks at the outburst epoch and decays exponentially at a characteristic time of 1~d as discussed in \S~\ref{sec:morphology}.

The size distribution is set to the interval of $\beta_\mathrm{rp,max}=1$ to an upper size limit constrained by the escape speed $v_\mathrm{esc}=\sqrt{2GM_\mathrm{N}/R_\mathrm{G}}$ where $M_\mathrm{N}=\frac{4}{3}\pi R_\mathrm{N}^3 \rho_\mathrm{N}$ is the total mass of the nucleus, $\rho_\mathrm{N}=500~\mathrm{kg \cdot m^{-3}}$ the bulk density of the nucleus, $R_\mathrm{N}=0.92$~km the effective nucleus radius \citep{Fernandez2013b}, and $R_\mathrm{G}=10R_\mathrm{N}$ the characteristic distance that gas drag become negligible \citep{Gombosi1986}. We only consider $\beta_\mathrm{rp,max}=1$ as (1) optical observations are most sensitive to $\beta_\mathrm{rp}\sim1$ (micron-sized) particles; (2) larger particles stay closer to the nucleus (as gravitational force dominates), models with $\beta_\mathrm{rp,max}\ll1$ are incompatible with the observations as they are not able to reproduce the obscured extended dust tails; and (3) complications arise for the dynamics of $\beta_\mathrm{rp,max}\gg1$ (submicron-sized) particles as there are also subjected to Lorentz forces.

We use the MERCURY6 package \citep{Chambers1999a} to integrate particles from the start epoch (i.e. the outburst epoch) to the observation epoch, using the 15th order RADAU integrator \citep{Everhart1985}. To accommodate the uncertainty in the exact epoch of the outburst, multiple outburst epochs, cued by the reports discussed in \S~\ref{sec:obs-reports}, are tested in the simulation. The production of simulated particles peaks at the assumed outburst epoch and decays exponentially afterwards, with a characteristic decay time of 1~d as found earlier in \S~\ref{sec:morphology}. Gravitational perturbations from the eight major planets (the Earth-Moon system is represented by a single mass at the barycenter of the two bodies), radiation pressure and Poynting-Robertson effect are included in the integration. 15P/Finlay's orbital elements are extracted from the JPL small body database elements K085/15 (\url{http://ssd.jpl.nasa.gov/sbdb.cgi}) as listed in Table~\ref{tbl:mdl-orb}.

\begin{figure*}
\includegraphics[width=\textwidth]{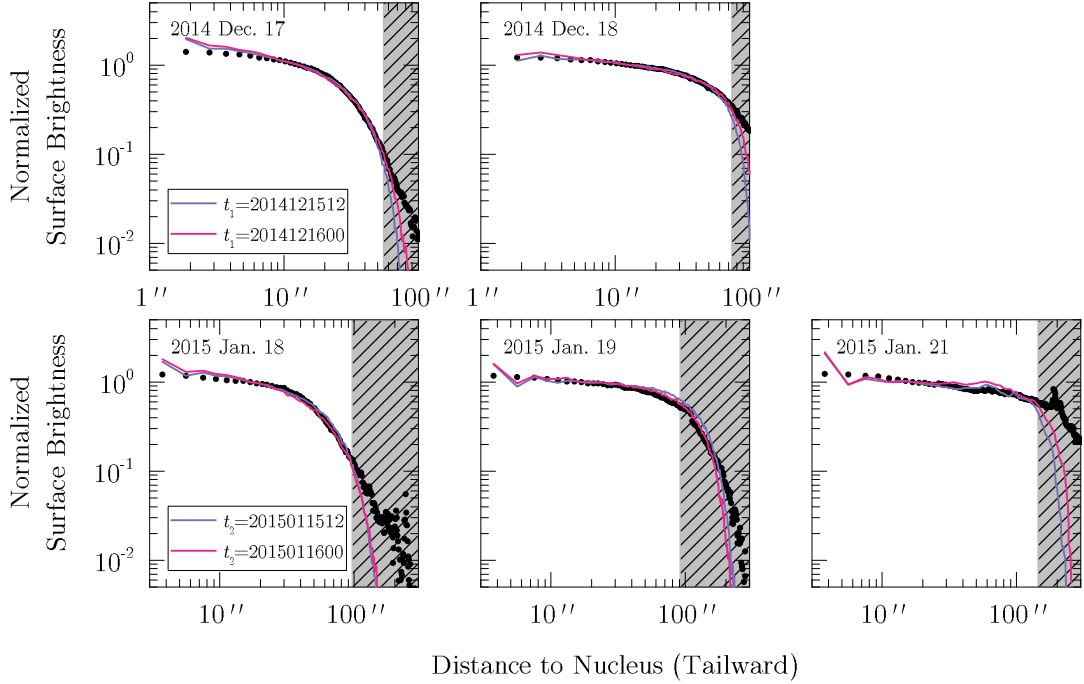}
\caption{Observed surface brightness profiles (scatter dots) and the best-fit dust models (color lines) for FRAM and Vicksburg observations. The assumed outburst epochs (see main text) are denoted as $t_1$ for the first outburst and $t_2$ for the second outburst. The regions that are dominated by submicron-sized particles are masked from modeling as described in the main text. For the profile on 2015 Jan. 19 additional region is masked due to the contamination of a background star.}
\label{fig:mdl}
\end{figure*}

The resulting modeled image is convolved with a 2-dimensional Gaussian function (with FWHM equal to the FWHM of the actual images) to mimic observational effects such as the instrumental point spread effect and atmospheric seeing. The modeled image is then collapsed into a 1-dimensional profile as what was done with the observations (\S~\ref{sec:obs-morphological}). Observed and modeled surface brightness profiles are normalized to 3~FWHMs beyond the nucleus along the Sun-comet axis. We mask out the region within 1~FWHM from the nucleus to avoid contamination of the signal from the nucleus. The region that is dominated by submicron-sized dust (i.e. the tailward region that is too far from the nucleus for $>1~\micron$ dust to reach) is also masked, as we are focused on $\micron$ to mm-sized dust. To evaluate the degree of similarity between the observed and the modeled profiles, we calculate the normalized error variance (NEV) as defined by

\begin{equation}
\label{eq:nev}
\mathrm{NEV} = \frac{1}{n} \sum_{i=1}^n \frac{\sqrt{(\mathbf{M}_i - \mathbf{O}_i)^2}}{\mathbf{O}_i}
\end{equation}

\noindent where $n$ is the number of pixels, $\mathbf{M}_i$ and $\mathbf{O}_i$ are the pixel brightness from the modeled and observed brightness profile respectively. We set the tolerance level of NEV to 10\%. The input parameters, test grids and best-fit results are tabulated in Table~\ref{tbl:mdl-orb}, ~\ref{tbl:mdl-parameter} and Figure~\ref{fig:mdl}.

It is encouraging that the best-fit models under the respective outburst epochs are largely consistent. We conclude that the characteristic ejection speed $V_0=300$~to~$650~\mathrm{m \cdot s^{-1}}$ for the ejecta of the first outburst, while $V_0=550$~to~$750~\mathrm{m \cdot s^{-1}}$ for the ejecta of the second outburst. The dust size index is at the range of $q\approx-3.5$. The ejection speed is comparable or is slightly larger than the one derived from the classic \citet{Whipple1950b} model (which gives $V_0 \sim 400~\mathrm{m \cdot s^{-1}}$ in our model), while the size index is comparable to the classic value, $q=-3.6$ \citep{Fulle2004}. It also appears that the characteristic ejection speed of the second outburst is higher than that of the first outburst, which seemingly support the idea that the second outburst was a more energetic event than the first one.

\section{Discussion}

\subsection{Nature of the Outburst}

The total mass of the dust emitted in the two outbursts are related to the effective scattering cross-section of the materials, $C_\mathrm{e}$, that can be calculated by

\begin{equation}
C_\mathrm{e} = \left( \frac{r_\mathrm{H}}{1~\mathrm{AU}} \right)^2 \frac{\pi \varDelta^2}{A_\lambda(\alpha)} 10^{0.4(m_{\odot, \lambda}-m_\lambda)}
\end{equation}

\noindent where $A_\lambda(\alpha)$, the phase angle corrected geometric albedo, is calculated using the compound Henyey-Greenstein model by \citet{Marcus2007b}, assuming $A_\lambda(0^\circ)=0.05$, and $m_{\odot, \lambda}$, $m_\lambda$ are the apparent magnitudes of the Sun and the comet. This yields $C_\mathrm{e}=7\times10^3~\mathrm{km^2}$ for the first outburst and $C_\mathrm{e}=2\times10^4~\mathrm{km^2}$ for the second outburst, using the photometric measurements in Figure~\ref{fig:report}. The total mass of the ejecta can then be calculated via $M_\mathrm{d}=\frac{4}{3} \rho_\mathrm{d} \bar{a}_\mathrm{d} C_\mathrm{e}$, where the mean dust size $\bar{a}_\mathrm{d}$ can be derived from the dust model discussed in \S~\ref{sec:kinematics}. Considering the variances among the best-fit models, we derive $M_\mathrm{d}=2$~to~$3\times10^5~\mathrm{kg}$ for the first outburst and $M_\mathrm{d}=4$~to~$5\times10^5~\mathrm{kg}$ for the second outburst (depending on the exact timing of the individual outburst), corresponding to less than $10^{-7}$ of the nucleus mass assuming a spherical nucleus.

With this mass, the specific energy of the two outbursts is calculated to be $0.3$~to~$2\times10^5~\mathrm{J \cdot kg^{-1}}$ using the speed component derived from the dust model. This value is comparable to the value derived for 17P/Holmes's 2007 outburst \citep[$\sim 10^5~\mathrm{J \cdot kg^{-1}}$, c.f.][]{Reach2010,Li2011}. For the case of 17P/Holmes, the large distance to the Sun at the time of its outburst, as well as the closeness of the derived specific energy to the specific energy of the amorphous ice to crystalline, are compatible with the idea that the comet's mega-outburst was triggered by the nergy released by the crystallization of amorphous ice. However, 15P/Finlay was much closer to the Sun at its two outbursts than 17P/Holmes at its 2007 outburst (1.0~AU versus 2.5~AU) such that solar heat may be sufficient to drive the outburst to some degree, hence we consider it difficult to assess the role of crystallization for 15P/Finlay's outburst at this stage.

\subsection{The Finlayid Puzzle Revisited}

15P/Finlay is puzzling in the sense that despite its occasional proximity to the Earth's orbit, the hypothetical Finlayid meteor shower has never been observed. This matter has been discussed in depth by \citet{1999MNRAS.310..168B}, who concluded that the perturbation of Jupiter has effectively dispersed the meteoroid stream, such that $\gtrsim99\%$ of the meteoroids released $\sim 20$~orbits ago would end up with distant \textit{nodal} passages ($>0.01$~AU) from the Earth's orbit. However, we think that this conclusion is unconvincing as the nodal plane approximation for Earth impact may not be valid for 15P/Finlay due to its shallow orbital plane ($i=6.8^\circ$). Additionally, new astrometric observations of 15P/Finlay in the last decade has reduced the uncertainty of the orbital elements by an order of magnitude; hence the issue of the long term evolution of the Finlayid meteoroid stream is worth revisiting.

\begin{figure*}
\includegraphics[width=\textwidth]{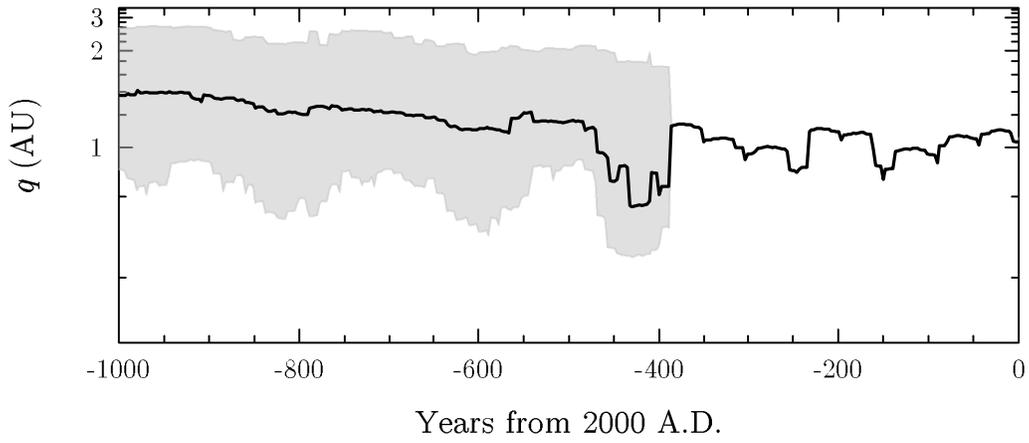}
\caption{Dynamical evolution of the perihelion distance of 100~clones of 15P/Finlay in the interval of 1000--2000~A.D.}
\label{fig:comet-dyn}
\end{figure*}

We first investigate the orbital stability of 15P/Finlay. This is done by generating 100~clones of 15P/Finlay using the orbital covariance matrix provided in JPL~K085/15, and integrating all of them $10^3$~yr backwards. The integration is performed with MERCURY6 using the Bulirsch-Stoer integrator. The evolution of the perihelion distance of all clones is shown in Figure~\ref{fig:comet-dyn}. It can be seen that the perihelion distance of the clones are highly compact until 1613~A.D., when a close encounter (miss distance of the order of 0.1~AU) between 15P/Finlay and Jupiter occurred. This implies that any backward meteoroid stream simulation will be physically meaningful only as long as the starting date is after 1613~AD.

\begin{figure}
\includegraphics[width=0.5\textwidth]{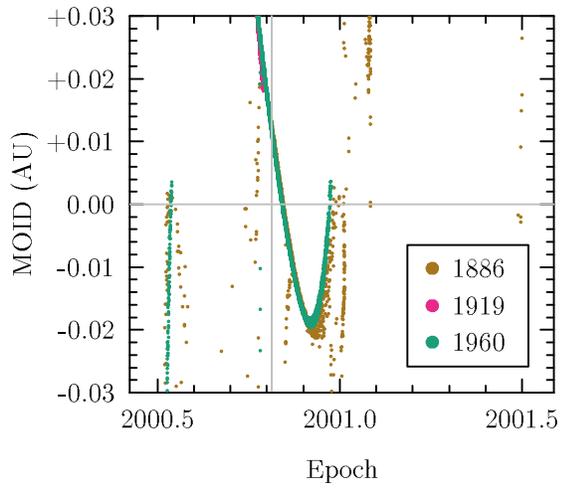}
\caption{The distribution of the dust trails released by 15P/Finlay during its 1886, 1909 and 1960 perihelion passages in 2001. Vertical gray line marks the time that the Earth passes the trails. It can be seen that the trails cross the Earth's orbit, suggestive of the possibility of a direct encounter to the Earth.}
\label{fig:met-2001}
\end{figure}

Next, we then simulate a total of 39,000 (a randomly chosen number) hypothetical particles released by 15P/Finlay during its 1886, 1909 and 1960 perihelion passage and examine their distribution in 2001, to directly compare to \citet{1999MNRAS.310..168B}'s simulations. The simulation is performed using the same collection of subroutines described in \S~\ref{sec:kinematics} except that the ejection model by \citet{Crifo1997a} is used and only $\beta=0.001$ (millimeter-sized) particles are simulated. The Minimum Orbit Intersection Distance vector $\overrightarrow{MOID}$ between the Earth and each meteoroid is calculated using the subroutine developed by \citet{Gronchi2005a} to assess the likelihood of an Earth encounter. The original MOID is defined as a scalar; here I define the direction of $\overrightarrow{MOID}$ to be the same as $r_\mathrm{D}-r_\mathrm{E}$, where $r_\mathrm{D}$, $r_\mathrm{E}$ are the heliocentric distance of the meteoroid and the Earth at the MOID point. We find as much as $\sim15\%$ of the particles stay within 0.01~AU from the Earth's orbit as of 2001, different from \citet{1999MNRAS.310..168B}'s finding. In addition, the dust trail is able to overlap with the Earth's orbit (Figure~\ref{fig:met-2001}), further supporting the idea that a significant number of particles released by 15P/Finlay 10--20 orbits ago may still have direct encounters with the Earth.

The background meteoroid flux originating from 15P/Finlay may be estimated in an order-of-magnitude manner. The absolute magnitude of 15P/Finlay is $\sim 100$~times brighter than low-activity comet 209P/LINEAR for which the meteoroid production capacity has been measured to be $10^{14}$ meteoroids per orbit \citep{Ye2016}. Hence, in 10~orbits, 15P/Finlay would generate $10^{17}$ meteoroids. Assuming the meteoroids distribute uniformly along the orbit with an orbital period of 5~yr, as well as a delivery efficiency of $10\%$ to the region $\pm 0.01$~AU from the Earth's orbit and a characteristic duration of meteor activity of 1~week, the flux can be calculate by $10^{17} \times 10\% \times 7~\mathrm{d} / 5~\mathrm{yr} \approx 0.1~\mathrm{km^{-2} \cdot hr^{-1}}$, which should be detectable by modern meteor survey systems.

To look for any undetected Finlayid activity, we conduct a ``cued'' search in the Canadian Meteor Orbit Radar (CMOR) database. CMOR is an interferometric radar array located near London, Canada operating at 29.85~MHz with a pulse repetition frequency of 532~Hz \citep[c.f.][]{Jones2005k, Brown2008x, Ye2013jq}. Since its commission in 2002, CMOR has measured 12 million meteoroid orbits, making it suitable for the search for weak meteor showers such as the Finlayids. We first calculate the characteristics of the hypothetical Finlayid radiant using the simulation results above, which yields $\lambda-\lambda_\odot=66^\circ\pm11^\circ$, $\beta=-18^\circ\pm9^\circ$ at Sun-centered ecliptic coordinates, and a geocentric speed of $13\pm3~\mathrm{km \cdot s^{-1}}$. We then combine 14 years of CMOR data into a stacked ``virtual'' year and look for any enhanced activity at the location of the theoretical radiant, using a wavelet-based search algorithm \citep{Brown2008x, Brown2010l} with probe sizes tuned to the expected radiant characteristics (radiant probe size $\sigma_\mathrm{rad}=10^\circ$, velocity probe size $\sigma_\mathrm{v}=3~\mathrm{km \cdot s^{-1}}$).

\begin{figure}
\includegraphics[width=0.5\textwidth]{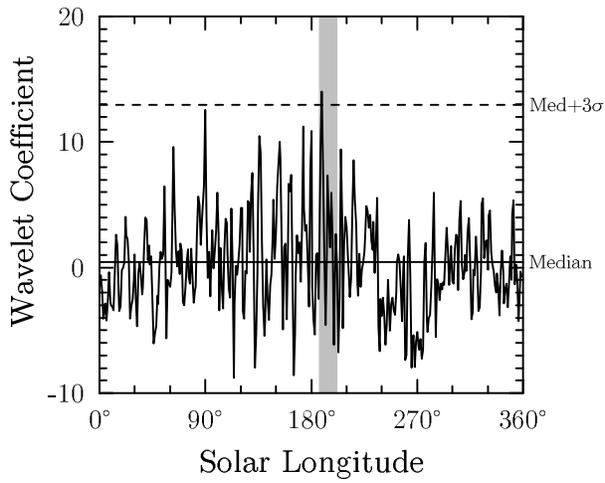}
\caption{The variation of the wavelet coefficient at the calculated Finlayid radiant $\lambda-\lambda_\odot=66^\circ$, $\beta=-25^\circ$, $v_\mathrm{G}=13~\mathrm{km \cdot s^{-1}}$ using the stacked ``virtual year'' CMOR data. The shaded area is the expected time window for Finlayid activity (solar longitude $\lambda_\odot \sim 193^\circ$).}
\label{fig:met-cmor}
\end{figure}

As shown in Figure~\ref{fig:met-cmor}, no significant enhancement can be found at the expected period of activity (solar longitude $\lambda_\odot \sim 210^\circ$). \citet{Ye2016} has calculated that the detection limit for the wavelet algorithm applied on CMOR is at the order of $0.01~\mathrm{km^{-2} \cdot hr^{-1}}$; however, CMOR sensitivity is also an order of magnitude less at a southerly radiant at $\delta=-40^\circ$ comparing to northerly radiants, so the shower flux limit is probably closer to $0.1~\mathrm{km^{-2} \cdot hr^{-1}}$. Hence, the existence and intensity (or derived upper-limit) of the Finlayid meteor shower is not definitive, but favors southern hemisphere meteor surveys \citep[e.g.][]{Younger2012,Janches2013}.

\subsection{The 2021 Earth Encounter of the 2014/2015 Outburst Ejecta}

The potential 2021 encounter with the 2014 trail from 15P/Finlay is of particular interest given the additional dust released from the two outbursts, as it offers an excellent opportunity to examine 15P/Finlay's ejecta. The encounter is studied by simply extending the numerical integration described in \S~\ref{sec:kinematics} to the year of 2021. Similar to the meteoroid trail model presented in \citet{Ye2016}, we assigned a \textit{space criterion} to select Earth-approaching meteoroids, defined by 

\begin{equation}
 \Delta X = v_\mathrm{rel} \times \Delta T
\end{equation}

where $v_\mathrm{rel}$ is the relative velocity between the meteoroid and the Earth and $\Delta T$ and $\delta T$ is called the \textit{time criterion} that is the characteristic duration of the event, typically $\Delta T=1$~d. Similar to the dust model discussed in \S~\ref{sec:kinematics}, we test two sets of outburst epochs that correspond to either the epochs of the last non-outburst report or the first positive report.

\begin{figure*}
\includegraphics[width=\textwidth]{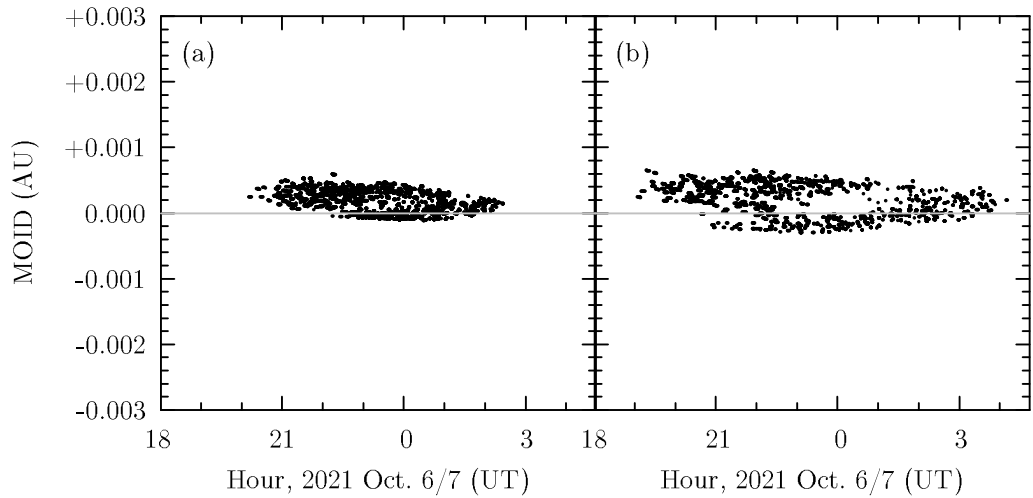}
\caption{Encounter of 15P/Finlay's 2014/2015 outburst ejecta in 2021 Oct. 6/7. Subfigure (a) corresponds to the simulation results assuming the earliest possible outburst epoch (2014 Dec. 15.4 UT for the first outburst, 2015 Jan. 15.5 UT for the second outburst), while (b) corresponds to the results assuming the latest possible outburst epoch (2014 Dec. 16.0 UT for the first outburst, 2015 Jan. 16.0 UT for the second outburst).}
\label{fig:met-2021}
\end{figure*}

The simulation result confirms the general findings by Maslov and Sato, that a direct encounter of the 2014/2015 meteoroid trail will occur on 2021 Oct. 6/7 (Table~\ref{tbl:met-2021}, Figure~\ref{fig:met-2021}) with a full-width-half-maximum (FWHM) of about 1~hour. The uncertainty in outburst epochs results in about 0.5~hour uncertainty in the peak time in 2021. The ejecta from the second outburst are calculated to arrive around 2021 Oct. 6 at 22~h UT, followed by those from the first outburst which are expected to arrive around 2021 Oct. 7 at 1~h UT. The radiant is at geocentric equatorial coordinates of $\alpha_\mathrm{G}=257^\circ$, $\delta_\mathrm{G}=-48^\circ$ or in the constellation of Ara, favoring the observers in the southern tip of Africa. As 15P/Finlay was $\sim20$ times more active during the two outbursts compared to its normal dust production level as indicated by Figure~\ref{fig:report}, the meteor activity may also be significantly stronger than previously expected. However, we also note that the range of the meteoroid sizes delivered to the Earth's vicinity seems to be concentrated at the order of $\beta \sim 0.001$, which translates to a visual magnitude of $+8$ \citep{Campbell-Brown2004c} considering the very low encounter speed. This indicates that the meteor activity in 2021 may only be visible to meteor radars and low-light video cameras.

\section{Summary}

We present an analysis of the two outbursts of the potentially comet-asteroid transition object, 15P/Finlay, at its 2014/2015 perihelion passage. These outbursts took place between 2014 Dec. 15.4--16.0 UT and 2015 Jan. 15.5--16.0 UT as constrained by ground-based and spacecraft observations. As seen in monitoring images, both outbursts produced a newly-formed dust shell that expands and fades in $\sim 1$ week.

The images from five observing nights (two for the first outburst, three for the second outburst) were studied using a Monte Carlo dust model, and yield a characteristic ejection speed of $V_0=300$~to~$650~\mathrm{m \cdot s^{-1}}$ for the ejecta of the first outburst and $V_0=550$~to~$750~\mathrm{m \cdot s^{-1}}$ for that of the second outburst, taking into account the uncertainty in the determination of outburst epoch. The dust size index is in the range of $q\approx-3.5$. We derive the mass of the ejecta to be $M_\mathrm{d}=2$~to~$3\times10^5~\mathrm{kg}$ for the first outburst and $M_\mathrm{d}=4$~to~$5\times10^5~\mathrm{kg}$ for the second outburst, corresponding to less than $10^{-7}$ of the nucleus mass. The specific energy of the two outbursts is calculated to be $0.3$~to~$2\times10^5~\mathrm{J \cdot kg^{-1}}$, comparable to the specific energy produced by the crystallization of amorphous ice, but does not prove the latter as the driving force for 15P/Finlay's outbursts.

We also revisited the long-standing puzzle of the non-detection of the Finlayids, the hypothetical meteor shower generated by 15P/Finlay, as well as the future possibility for meteor activity generated by the 2014/2015 outbursts. We find the efficiency of meteoroid delivery to the Earth's orbit is $\sim 10$~times higher than previously reported by \citet{1999MNRAS.310..168B}. Assuming 15P/Finlay's recent (last $\sim20$~orbits) activity is comparable to its contemporary level, the meteoroid flux of the Finlayids should be high enough to be detected by modern meteor surveys. However, a cued search with the 12 million meteor orbits gathered by the Canadian Meteor Orbit Radar over the past 13 years does not reveal any positive detection. The encounter with the 2014/2015 outburst ejecta may provide an answer to the Finlayid puzzle, as the Earth is expected to pass though the ejecta trails directly around 2021 Oct. 6 at 22~h UT to Oct. 7 at 1~h UT, with a chance for some significant meteor activity in the video or radio range. The timing and the southerly radiant in the constellation of Ara will favor observers in the southern tip of Africa.

The recent outburst episode of 15P/Finlay seems to suggest that the comet, originally thought to be quiet and largely inactive, does possess the ability for significant activity. Whether the recent outbursts are the overtures of a resurrection of the comet or a finale of its career remains to be seen. Cometary observations in the forthcoming perihelion passage in July 2021, as well as observations during the potential meteor outburst, will likely provide more information.

\acknowledgments

We thank an anonymous referee for his/her comments. We also thank Paul Wiegert for permission to use his computational resource, Margaret Campbell-Brown for her meteoroid ablation code, Zbigniew Krzeminski, Jason Gill, Robert Weryk and Daniel Wong for helping with CMOR operations. The operation of the robotic telescope FRAM is supported by the EU grant GLORIA (No. 283783 in FP7-Capacities program) and by the grant of the Ministry of Education of the Czech Republic (MSMT-CR LG13007). Funding support from the NASA Meteoroid Environment Office (cooperative agreement NNX11AB76A) for CMOR operations is gratefully acknowledged. This work would not be possible without the inputs from the amateur community and their contributions are gratefully acknowledged.


\bibliographystyle{apj}

\clearpage

\clearpage

\clearpage

\clearpage

\clearpage

\clearpage

\clearpage

\clearpage

\clearpage

\begin{table*}
\centering
\caption{Summary of the imaging observations.}
\begin{tabular}{lclccccc}
\hline
Date & Time\tablenotemark{a} & Facility\tablenotemark{b} & Total Exposure & Filter & $r_\mathrm{H}$ & $\varDelta$ & Plane Angle \\
     & (UT) &                 & (min.)           & & (AU) & (AU) & \\ 
\hline
2014 Dec. 17 & 00:57 & FRAM & 8 & $R_\mathrm{C}$ & 0.987 & 1.472 & $4.3^\circ$ \\
2014 Dec. 18 & 00:57 & FRAM & 8 & $R_\mathrm{C}$ & 0.985 & 1.467 & $4.3^\circ$ \\
2014 Dec. 18 & 12:01 & Xingming & 10 & Unfiltered & 0.984 & 1.465 & $4.3^\circ$ \\
2014 Dec. 22 & 12:10 & Xingming & 24 & Unfiltered & 0.978 & 1.446 & $4.5^\circ$ \\
2014 Dec. 23 & 12:18 & Xingming & 51 & Unfiltered & 0.977 & 1.442 & $4.5^\circ$ \\
2014 Dec. 30 & 12:20 & Xingming & 26 & Unfiltered & 0.977 & 1.416 & $4.7^\circ$ \\
\hline
2015 Jan. 17 & 12:43 & Xingming & 83 & Unfiltered & 1.025 & 1.394 & $4.7^\circ$ \\
2015 Jan. 18 & 00:30 & Vicksburg & 3 & $R_\mathrm{C}$ & 1.023 & 1.393 & $4.7^\circ$ \\
2015 Jan. 19 & 00:18 & Vicksburg & 6 & $R_\mathrm{C}$ & 1.028 & 1.395 & $4.6^\circ$ \\
2015 Jan. 19 & 13:41 & Xingming & 16 & Unfiltered & 1.035 & 1.396 & $4.6^\circ$ \\
2015 Jan. 20 & 12:53 & Xingming & 91 & Unfiltered & 1.040 & 1.397 & $4.6^\circ$ \\
2015 Jan. 21 & 00:44 & Vicksburg & 6 & $R_\mathrm{C}$ & 1.042 & 1.398 & $4.7^\circ$ \\
2015 Jan. 21 & 13:13 & Xingming & 88 & Unfiltered & 1.045 & 1.399 & $4.6^\circ$ \\
2015 Jan. 23 & 13:29 & Xingming & 17 & Unfiltered & 1.055 & 1.403 & $4.6^\circ$ \\
2015 Jan. 27 & 13:19 & Xingming & 88 & Unfiltered & 1.078 & 1.415 & $4.3^\circ$ \\
2015 Jan. 29 & 13:32 & Xingming & 68 & Unfiltered & 1.091 & 1.422 & $4.2^\circ$ \\
\hline
\end{tabular}
\label{tbl:obs-summary}
\end{table*}

\clearpage

\begin{table*}
\centering
\caption{General parameters for the dust model. The orbital elements are quoted from the JPL elements K085/15. The nucleus radius is reported by \citet{Fernandez2013b}.}
\begin{tabular}{ll}
\hline
Parameter & Value \\
\hline
Semimajor axis $a$ & 3.48762~AU \\
Eccentricity $e$ & 0.72017 \\
Inclination $i$ & $6.79902^\circ$ \\
Longitude of the ascending node $\Omega$ & $13.77506^\circ$ \\
Argument of perihelion $\omega$ & $347.55924^\circ$ \\
Epoch of perihelion passage $t_\mathrm{p}$ & 2014 Dec. 27.05599 UT \\
Total magnitude $H$ & 10.7 \\
Non-grav. radial acceleration parameter $A_1$ & $4.075\times10^{-9}~\mathrm{AU/d^2}$ \\
Non-grav. transverse acceleration parameter $A_2$ & $5.744\times10^{-11}~\mathrm{AU/d^2}$ \\
Nucleus radius $r_\mathrm{c}$ & 0.9~km \\
Nucleus bulk density $\rho_\mathrm{c}$ & $500~\mathrm{kg \cdot m^{-3}}$ (assumed) \\
Dust bulk density $\rho_\mathrm{d}$ & $1000~\mathrm{kg \cdot m^{-3}}$ (assumed) \\
Minimum dust size $\beta_\mathrm{rp,max}$ & $1.0$ \\
\hline
\end{tabular}
\label{tbl:mdl-orb}
\end{table*}

\clearpage

\begin{table*}
\centering
\caption{Best-fit dust models for the FRAM and Vicksburg observations.}
\begin{tabular}{lllcc}
\hline
 & Outburst Epoch & Observation Epoch & $V_0$ & $q$ \\
 & (UT) & (UT) & $\mathrm{m \cdot s^{-1}}$ & \\
\hline
Test grids & - & - & 100 to 1200 & -5.4 to -2.0 \\
 & & & in steps of 20 & in steps of 0.1 \\
\hline
1$^\mathrm{st}$ Outburst & 2014 Dec. 15, 10~h & 2014 Dec. 17 & $320 \pm 10$ & $-3.7 \pm 0.2$ \\
 & .. & 2014 Dec. 18 & $320 \pm 20$ & $-3.0 \pm 0.3$ \\
 & 2014 Dec. 16, 0~h & 2014 Dec. 17 & $640 \pm 30$ & $-4.0 \pm 0.6$ \\
 & .. & 2014 Dec. 18 & $670 \pm 90$ & $-3.7 \pm 0.5$ \\
\hline
2$^\mathrm{nd}$ Outburst & 2015 Jan. 15, 12~h & 2015 Jan. 18 & $540 \pm 40$ & $-3.6 \pm 0.6$ \\
 & .. & 2015 Jan. 19 & $590 \pm 120$ & $-3.4 \pm 0.5$ \\
 & .. & 2015 Jan. 21 & $570 \pm 30$ & $-3.6 \pm 0.6$ \\
 & 2015 Jan. 16, 0~h & 2015 Jan. 18 & $780 \pm 30$ & $-3.8 \pm 0.5$ \\
 & .. & 2015 Jan. 19 & $670 \pm 100$ & $-3.6 \pm 0.5$ \\
 & .. & 2015 Jan. 21 & $750 \pm 40$ & $-3.4 \pm 0.4$ \\
\hline
\end{tabular}
\label{tbl:mdl-parameter}
\end{table*}

\clearpage

\begin{table*}
\centering
\caption{Predictions of the 2021 encounter of 15P/Finlay's 2014 meteoroid trails.}
\begin{tabular}{llcccc}
\hline
 & Peak Time & Radiant & $v_\mathrm{g}$ & Note \\
 & (UT) & $\alpha_\mathrm{g}, \delta_\mathrm{g}$ & $\mathrm{km \cdot s^{-1}}$ & $\mathrm{hr^{-1}}$ \\
\hline
Maslov\tablenotemark{a} & 2021 Oct. 7, 1:19 & $255.8^\circ, -48.3^\circ$ & 10.7 & ZHR 5--50 \\
Sato\tablenotemark{b} & 2021 Oct. 7, 1:10 & $255.7^\circ, -48.4^\circ$ & 10.7 & - \\
Vaubaillon\tablenotemark{c} & - & - & - & No encounter \\
This work & 2021 Oct. 7, 0:34--1:09\tablenotemark{d} & $255.6^\circ, -48.4^\circ$ & $10.7$ & Ejecta from the first outburst \\
  & 2021 Oct. 6, 21:59--22:33\tablenotemark{e} & $256.3^\circ, -48.5^\circ$ & $10.7$ & Ejecta from the second outburst \\
\hline
\end{tabular}
\tablenotetext{a}{\url{http://feraj.narod.ru/Radiants/Predictions/1901-2100eng/Finlayids1901-2100predeng.html}, retrieved 2015 Mar. 4.}
\tablenotetext{b}{\url{https://groups.yahoo.com/neo/groups/meteorobs/conversations/messages/44030}, retrieved 2015 Mar. 4.}
\tablenotetext{c}{\url{https://groups.yahoo.com/neo/groups/meteorobs/conversations/messages/44035}, retrieved 2015 Mar. 4.}
\tablenotetext{d}{The peak time of 0:34 corresponds to the assumed outburst epoch of 2014 Dec. 15.4 UT, while 1:09 corresponds to the assumed outburst epoch of 2014 Dec. 16.0 UT.}
\tablenotetext{e}{The peak time of 21:59 corresponds to the assumed outburst epoch of 2015 Jan. 16.0 UT, while 22:33 corresponds to the assumed outburst epoch of 2015 Jan. 15.5 UT.}
\label{tbl:met-2021}
\end{table*}

\end{CJK*}
\end{document}